*Original Article*

# Almansi-type boundary conditions for electric potential inducing flexure in linear piezoelectric beams

F. dell'Isola, L. Rosa

Dipartimento Ingegneria Strutturale e Geotecnica, Università di Roma "La Sapienza" Via Eudossiana n. 18, I-00184 Roma Italia

Using the recent results found in [1, 2] we prove that it is possible to induce flexur in linear piezoelectric beams by means of quadratic Almansi type boundary conditions for the electric potential. Beams constituted by transversely isotropic piezoelectric materials whose symmetry axis is parallel to the axis of the beam are considered. Our choice of boundary conditions for the electric potential has been suggested by the results found in [1, 3]. An explicit expression of material parameters that influenc flexur is given in terms of piezoelectric moduli.

## 1 Introduction

In a recent paper Batra and Yang [2] demonstrated the validity of the Saint-Venant (SV) principle for linear piezoelectricity. Their results open the related question of findin the solutions of the SV-Almansi problem for a SV cylinder constituted by linear Transversally Isotropic Piezoelectric (TIP) material. We assume that the material symmetry axis is parallel to the axis of the cylinder. However it is implicitly implied by the results found in [3] that for homogeneous electrical and mechanic boundary conditions – on the lateral wall of the cylinder – one should expect a very weak coupling between mechanical and electrical phenomena.

Under the Clebsch-SV hypothesis expressed by Eq. (29), we prove that, in absence of an external electric fiel orthogonal to the axis of the cylinder (z-axis), the only electro-mechanical coupling occurs under pure extension and compression. On the other hand, when the boundary value of the electric potential varies along the z-axis, very interesting coupling effects can arise. Indeed in the case of Almansi-type boundary conditions -quadratic in z- we fin that coupling of electrical and mechanical effects occurs in flexure

The considered prismatic beam in the reference configuratio occupies a volume $C$ which is the Cartesian product of a plane section $C_\pi$ times a straight line segment $l := [-L, L]$. We will call *lateral wall* the Cartesian product $\partial C_\pi \times l$ and *bases* the sets $C_\pi \times \{-L\}$ and $C_\pi \times \{L\}$. In this paper we assume that this space is fille by a TIP material, and then formulate the Clebsch-SV hypothesis on stress in order to apply the semi-inverse method of SV to solve an Almansi-type problem. In particular we prove that it is possible to control the SV flexur and Poisson effect in SV compression and extension.

The balance equations describing the local equilibrium for piezoelectricity can be found, for instance, in [2, 4, 8, 9]

$$\text{DIV}\,\mathbf{T} = 0, \quad \text{DIV}\,\mathbf{D} = 0, \tag{1}$$

where $\mathbf{T}$ is the Cauchy stress tensor and $\mathbf{D}$ is the electric displacement vector.

In the Almansi-SV problem suitable boundary conditions on the lateral wall of the prismatic beam must be added to these equations describing the nature of mechanical and electrical interactions of the bar with the external world. We limit our analysis to the case of unloaded lateral walls ($\mathbf{n} \perp \partial C_\pi \times l$):

$$\mathbf{Tn}|_{\partial C_\pi \times l} = \mathbf{0}. \tag{2}$$

Moreover we will consider two kind of electrostatic interactions through the lateral wall:

i) the case in which the cylinder is in contact with a medium with low permittivity so that outside the prismatic bar the normal component of the electric induction vector vanishes [2]

$$\mathbf{D} \cdot \mathbf{n}|_{\partial C_\pi \times l} = 0, \tag{3}$$

ii) the case in which conductors at an assigned potential ($z$ is the coordinate along the axis of $C$) envelop the cylinder

$$\varphi|_{\partial C_\pi \times l} = \varphi_0(\mathbf{r}_\pi) + z\varphi_1(\mathbf{r}_\pi) + z^2 \varphi_2(\mathbf{r}_\pi), \qquad \mathbf{r}_\pi \in \partial C_\pi; \tag{4}$$

this is an Almansi-type boundary condition (see [5]).

We do not specify the boundary conditions on the bases. Indeed, up to some scalar constants they will be determined. The solution to the SV problem for linear elastic prismatic beams depends on six scalar constants (see for instance [10]) which are in one-to-one correspondence with the resultant force and torque applied on the bases. We will show that for a linear piezoelectric SV cylinder some extra constants are needed to determine the flux of the electric displacement vector through the bases.

The geometry of the problem allows for a very useful decomposition of the 3-d displacement vector field into a direct sum of vectors parallel to the axis $l$ and vectors lying in the plane section $C_\pi$ orthogonal to $l$. Following [11] we decompose all the tensor algebra and tensor differential operators correspondingly. In this way the calculations are substantially simplified in comparison to the original Almansi-SV problem [6, 10].

We prove that when boundary condition (3) applies it is possible to electrically control the Poisson effect. For boundary condition (4) the elliptic problem for the displacement field is coupled with the elliptic problem determining the electric potential. Postponing the study of these problems to further investigations we prove that for a particular choice of boundary condition (4) and for SV cylinders with rectangular cross sections, flexural displacement ("flexion inégale" following Saint Venant) can be controlled by means of the applied potential difference.

## 2 Constitutive equations

Because of the decomposition of the reference placement of the body as $C_\pi \times l$ we can identify a generic point in the beam by means of the pair $(\mathbf{r}_\pi, z)$ with $\mathbf{r}_\pi \in C_\pi$ and $z$ the coordinate along $l$. We decompose vectors and tensors as follows:

$$\mathbf{r} = z\mathbf{e}_z + \mathbf{r}_\pi, \quad \mathbf{u} = u_z \mathbf{e}_z + \mathbf{u}_\pi, \quad \mathbf{E} = E_z \mathbf{e}_z + \mathbf{E}_\pi, \tag{5}$$

$$\mathbf{D} = D_z \mathbf{e}_z + \mathbf{D}_\pi, \quad \Sigma = \hat{\Sigma} + \zeta \otimes \mathbf{e}_z + \mathbf{e}_z \otimes \zeta + \eta \mathbf{e}_z \otimes \mathbf{e}_z, \tag{6}$$

where $\mathbf{u}$ is the displacement vector, $\mathbf{E}$ the electric field $\mathbf{D}$ the electric displacement vector, $\mathbf{e}_z$ the unit vector in the direction of the symmetry axes $l$, $\pi$ denotes the projection onto $C_\pi$, $\mathbf{r}_\pi$ is the position vector from an arbitrary origin $o \in C_\pi$, $\hat{\Sigma}$ and $\zeta$ are respectively a second order tensor field and a vector field in $C_\pi$ and $\eta$ is a scalar field In the following (see constitutive equations) we will need the decomposition of transversely-isotropic tensors of rank two, three and four. Let $\mathbf{e}_z$ be the axes of symmetry, then we get [1, 3]:

$$\mathbf{C} = 2\mu \mathbf{I}_\pi \square \mathbf{I}_\pi + \lambda \mathbf{I}_\pi \otimes \mathbf{I}_\pi + \alpha_1 (\mathbf{P} \square \mathbf{I} + \mathbf{I} \square \mathbf{P}) +$$
$$+ \alpha_2 (\mathbf{P} \otimes \mathbf{I} + \mathbf{I} \otimes \mathbf{P}) + 2\alpha_3 \mathbf{P} \otimes \mathbf{P}, \tag{7}$$

$$\Xi = \beta_1 \mathbf{I}_\pi \otimes \mathbf{e}_z + \beta_2 (\mathbf{I}_\pi \square \mathbf{e}_z + \mathbf{e}_z \otimes \mathbf{I}_\pi) + \beta_3 \mathbf{P} \otimes \mathbf{e}_z, \tag{8}$$

$$\Sigma^d = \gamma_1 \mathbf{I}_\pi + \gamma_2 \mathbf{P}, \tag{9}$$

where $\mathbf{I}$ is the identity operator in $R^3$, $\mathbf{I}_\pi$ is the identity operator in $C_\pi$ and $\mathbf{P} = \mathbf{e}_z \otimes \mathbf{e}_z$ is the projector along the symmetry axes $l$. Furthermore, $\mu$, $\lambda$, $\alpha_i$, $\beta_i (i = 1, 2, 3)$, $\gamma_1$ and $\gamma_2$ are scalar coefficients The action of the previous tensors are determined as follows (using the following convention: $\mathbf{C} \cdot \mathbf{a} = \sum_j C_{ij} a_j$, $(\mathbf{C} \cdot \mathbf{B})_{ik} = \sum_j C_{ij} B_{jk}$, $\mathbf{C} : \mathbf{B} = \sum_{ij} C_{ij} B_{ji}$,):

$$(\mathbf{A} \square \mathbf{B}) : \mathbf{C} = \mathbf{A} \cdot \mathbf{C} \cdot \mathbf{B}^T \quad , \quad (\mathbf{A} \otimes \mathbf{B}) : \mathbf{C} = \mathbf{A}(\mathbf{C} : \mathbf{B}), \tag{10}$$

$$(\mathbf{A} \square \mathbf{a}) : \mathbf{B} = \mathbf{A} \cdot (\mathbf{B}^T \cdot \mathbf{a}) \quad , \quad (\mathbf{A} \otimes \mathbf{a}) : \mathbf{B} = \mathbf{A} \cdot (\mathbf{B} \cdot \mathbf{a}), \tag{11}$$

$$(\mathbf{A} \square \mathbf{a}) \mathbf{b} = (\mathbf{A} \cdot \mathbf{b}) \otimes \mathbf{a} \quad , \quad (\mathbf{A} \otimes \mathbf{a}) \mathbf{b} = \mathbf{A}(\mathbf{a} \cdot \mathbf{b}). \tag{12}$$

Having these in mind we may defin the following block matrix representations $\begin{pmatrix} 2 \times 2 & 2 \times 1 \\ 1 \times 2 & 1 \times 1 \end{pmatrix}$:

$$\mathbf{T} := \left( \begin{array}{c|c} \hat{\mathbf{T}} & \tau \\ \hline \tau^T & \sigma \end{array} \right), \tag{13}$$

$$\Sigma := \left( \begin{array}{c|c} \hat{\Sigma} & \zeta \\ \hline \zeta^T & \eta \end{array} \right), \tag{14}$$

$$\mathbf{C} : \Sigma = \left( \begin{array}{c|c} 2\mu\hat{\Sigma} + [\lambda tr(\hat{\Sigma}) + \alpha_2\eta]\mathbf{I}_\pi & \alpha_1\zeta \\ \hline \alpha_1\zeta^T & \alpha_2 tr(\hat{\Sigma}) + \alpha\eta \end{array} \right), \tag{15}$$

$$\Xi : \Sigma = \left( \begin{array}{c} (\beta_1 + \beta_2)\zeta \\ \beta_2 tr(\hat{\Sigma}) + \beta_3\eta \end{array} \right), \tag{16}$$

$$\Xi : \mathbf{E} = \left( \begin{array}{c|c} \beta_1 E_z \mathbf{I}_\pi & \beta_2 \mathbf{E}_\pi \\ \hline \beta_2 \mathbf{E}_\pi^T & \beta_3 E_z \end{array} \right), \tag{17}$$

$$\Sigma^d . \mathbf{E} = \left( \begin{array}{c} \gamma_1 \mathbf{E}_\pi \\ \gamma_2 E_z \end{array} \right), \tag{18}$$

where $\alpha = 2(\alpha_1 + \alpha_2 + \alpha_3)$.

TIP materials obey the following constitutive equations [1–3]:

$$\mathbf{T} = \mathbf{C} : \Sigma + \Xi : \mathbf{E}, \tag{19}$$

$$\mathbf{D} = \Sigma^d : \mathbf{E} - \Xi : \Sigma. \tag{20}$$

Thus we fin

$$\mathbf{T} = \left( \begin{array}{c|c} 2\mu\hat{\Sigma} + [\lambda tr(\hat{\Sigma}) + \alpha_2\eta + \beta_1 E_z]\mathbf{I}_\pi & \alpha_1\zeta + \beta_2 \mathbf{E}_\pi \\ \hline \alpha_1\zeta^T + \beta_2 \mathbf{E}_\pi^T & \alpha_2 tr(\hat{\Sigma}) + \alpha\eta + \beta_3 E_z \end{array} \right), \tag{21}$$

$$\left( \begin{array}{c} \mathbf{D}_\pi \\ D_z \end{array} \right) = \left( \begin{array}{c} \gamma_1 \mathbf{E}_\pi - (\beta_1 + \beta_2)\zeta \\ \gamma_2 E_z - \beta_2 tr(\hat{\Sigma}) - \beta_3\eta \end{array} \right). \tag{22}$$

For what concerns the gradient operator we have the following decomposition

$$\text{GRAD}(\varphi) = \varphi' \mathbf{e}_z + grad(\varphi), \tag{23}$$

$$\text{GRAD}(\mathbf{u}) = grad(\mathbf{u}_\pi) + grad(u_z) \otimes \mathbf{e}_z + \mathbf{u}'_\pi \otimes \mathbf{e}_z + u'_z \mathbf{e}_z \otimes \mathbf{e}_z \tag{24}$$

(here and in the following we write GRAD (DIV) for the gradient (divergence) operator that acts in $R^3$, $grad$ ($div$) for the one that acts in $C_\pi$ and, finally the ' stands for $\partial_z$. Moreover we will denote with $\Delta_\pi$ the restriction of Laplacian operator $\Delta$ to $C_\pi$).

We assume that the electric fiel $\mathbf{E}$ and the infinitesima deformation tensor $\Sigma$ are the gradient of a potential $\varphi$ and of the displacement vector $\mathbf{u}$, respectively,

$$\mathbf{E} = (grad\,\varphi, \varphi'),$$

$$\Sigma = \text{Sym}(\text{GRAD}(\mathbf{u})) = \left( \begin{array}{c|c} \text{Sym}(grad(\mathbf{u}_\pi)) & (\mathbf{u}'_\pi + grad(u_z))/2 \\ \hline (\mathbf{u}'_\pi + grad(u_z))/2 & u'_z \end{array} \right). \tag{25}$$

We will also need the following compatibility theorem (see for instance [12]):

- Let $f$ be a scalar fiel on $C_\pi$

$$\exists \mathbf{u} / \begin{cases} f \mathbf{I} = \text{Sym}\,grad(\mathbf{u}) \\ g* = \text{Skw}\,grad(\mathbf{u}) \end{cases} \Rightarrow \Delta f = 0, \; grad(g) = *grad(f). \tag{26}$$

Here, * denotes the (counter clockwise) $\pi/2$-rotation operator in the plane including $C_\pi$. Theorem (26) states that if a spherical tensor fiel is the symmetric part of the gradient of a vector field then its Laplacian is vanishing and can be uniquely completed by a skew part which is its rotation gradient.

## 3 Saint-Venant problem

In this section we derive the elliptic problems that characterize the deformation of piezoelectric SV beams under the Clebsch Saint-Venant hypothesis.

From $\text{\scriptsize DIV}\,\mathbf{D} = 0$ and with Eq.20 we may deduce

$$\gamma_1 \Delta_\pi \varphi - (\beta_1 + \beta_2) div(\frac{\mathbf{u}'_\pi}{2}) - (\beta_1 + \beta_2)\frac{div(grad(u_z))}{2} + A_2\varphi'' + B_2 u_z'' = 0 \tag{27}$$

with

$$A_2 = \gamma_2 + \beta_2\beta_1/(\mu+\lambda), \ B_2 = \beta_2\alpha_2/(\mu+\lambda) - \beta_3.$$

Denoting

$$\widetilde{\varphi} := \gamma_1 \varphi - \frac{1}{2}(\beta_1+\beta_2)u_z$$

we obtain

$$\Delta_\pi \widetilde{\varphi} - (\beta_1+\beta_2)div(\frac{\mathbf{u}'_\pi}{2}) + A_2\varphi'' + B_2 u_z'' = 0. \tag{28}$$

Now, we introduce the Clebsch Saint-Venant hypothesis

$$\hat{\mathbf{T}} = \mathbf{0}. \tag{29}$$

With it and by means of (21) one may deduce

$$tr\left(\hat{\Sigma}\right) = -\frac{\alpha_2}{\mu+\lambda}\eta - \frac{\beta_1}{\mu+\lambda}E_z, \tag{30}$$

$$\sigma = A_1\eta + B_1 E_z, \tag{31}$$

$$\tau = \alpha_1 \zeta + \beta_2 \mathbf{E}_\pi, \tag{32}$$

$$-2(\mu+\lambda)\hat{\Sigma} = [\alpha_2\eta + \beta_1 E_z]\mathbf{I}_\pi \tag{33}$$

in which $A_1 = 2(\alpha_1+\alpha_2+\alpha_3) - \alpha_2^2/(\mu+\lambda), \ B_1 = \beta_3 - \beta_1\alpha_2/(\mu+\lambda)$.

From $\text{\scriptsize DIV}\,\mathbf{T} = 0$ we fin

$$\sigma' + div\tau = 0, \ \tau' = \mathbf{0} \implies \sigma'' = 0. \tag{34}$$

Consequently, from (31)

$$E_z'' = -\frac{A_1}{B_1}\eta'' \implies \varphi''' = -\frac{A_1}{B_1}u_z''' \tag{35}$$

and from (25, 33)

$$\hat{\Sigma}'' = A_4\eta''\mathbf{I}_\pi \implies Sym(grad(\mathbf{u}''_\pi)) = A_4\eta''\mathbf{I}_\pi. \tag{36}$$

where

$$A_4 := -1/[2(\mu+\lambda)]\left[\alpha_2 - \beta_1\frac{A_1}{B_1}\right]$$

From $\tau' = \mathbf{0}$, by using (32) and (25), there follows

$$\alpha_1 \frac{\mathbf{u}''_\pi}{2} + grad\left(\frac{\alpha_1}{2}u'_z + \beta_2\varphi'\right) = \mathbf{0} \tag{37}$$

and, if $\tilde{u}_z := \alpha_1 u_z + 2\beta_2\varphi$

$$\alpha_1 grad(\mathbf{u}''_\pi) + grad(grad(\tilde{u}'_z)) = \mathbf{0} \implies Skw(grad(\mathbf{u}''_\pi)) = \mathbf{0}. \tag{38}$$

Consequently, by (36) and (38)

$$\alpha_1 A_4 \eta'' \mathbf{I}_\pi + grad(grad(\tilde{u}'_z)) = \mathbf{0}. \tag{39}$$

Equations $(36)_2$, $(38)_2$ and $(26)$ imply that $\eta''(\mathbf{r}_\pi, z)$ is a function of $z$ only.

Let $h(z) = -\alpha_1 A_4 \eta''(z)$; if we calculate the trace of (39), differentiate (28) with respect to $z$ and substitute the expression for $div(\mathbf{u}''_\pi)$ as obtained by $(36)_2$, then we obtain

$$\Delta_\pi \tilde{u}'_z = 2h(z), \tag{40}$$

$$\Delta_\pi \tilde{\varphi}' = (A_4(\beta_1 + \beta_2) + \frac{A_2 A_1}{B_1} - B_2)h(z). \tag{41}$$

Applying, next, condition (26) to $(36)_2$ and $(38)_2$ yields

$$\alpha_2 \Delta_\pi u'_z + \beta_1 \Delta_\pi \varphi' = 0. \tag{42}$$

In view of the definition of $\tilde{u}_z$ and $\tilde{\varphi}$ and by using (40) and (41), the last equation becomes

$$[\alpha_2 F_1 + \beta_1 G_1] h(z) = 0 \tag{43}$$

with

$$F_1 := \frac{2\gamma_1 - 2\beta_2(A_4(\beta_1 + \beta_2) + \frac{A_2 A_1}{B_1} - B_2)}{\gamma_1 \alpha_1 + \beta_2(\beta_1 + \beta_2)},$$

$$G_1 := \frac{\beta_1 + \beta_2 + \alpha_1(A_4(\beta_1 + \beta_2) + \frac{A_2 A_1}{B_1} - B_2)}{\gamma_1 \alpha_1 + \beta_2(\beta_1 + \beta_2)}.$$

If the material modulus (in square brackets) appearing in (43) is non-vanishing, then (43) requires that

$$h(z) = 0. \tag{44}$$

This will be assumed in the sequel so that, because of (35), we have

$$u'''_z = 0, \; \varphi''' = 0 \Rightarrow \tilde{u}'''_z = 0, \; \tilde{\varphi}''' = 0. \tag{45}$$

Thus we have found

$$grad(grad(\tilde{u}'_z)) = 0 \tag{46}$$

whose general integral is

$$\tilde{u}'_z(\mathbf{r}_\pi, z) = \mathbf{v}(z) \cdot \mathbf{r}_\pi + b(z); \tag{47}$$

differentiating this twice with respect to z we fin

$$\tilde{u}'''_z = \mathbf{v}''(z) \cdot \mathbf{r}_\pi + b''(z) = 0, \tag{48}$$

where $(45)_3$ has also been used. This is an $\mathbf{r}_\pi$-polynomial, and so we must set

$$\mathbf{v}''(z) = \mathbf{0}, \quad b''(z) = 0, \tag{49}$$

from which

$$\mathbf{v}(z) = \mathbf{v}_1 + \mathbf{v}_2 z, \quad b(z) = b_1 + b_2 z. \tag{50}$$

Again with the aid of (45) we get

$$u_z = u_z^0(\mathbf{r}_\pi) + u_z^1(\mathbf{r}_\pi)z + u_z^2(\mathbf{r}_\pi)\frac{z^2}{2}, \quad \tilde{u}_z = \tilde{u}_z^0(\mathbf{r}_\pi) + \tilde{u}_z^1(\mathbf{r}_\pi)z + \tilde{u}_z^2(\mathbf{r}_\pi)\frac{z^2}{2}, \tag{51}$$

$$\varphi = \varphi_0(\mathbf{r}_\pi) + \varphi_1(\mathbf{r}_\pi)z + \varphi_2(\mathbf{r}_\pi)\frac{z^2}{2}, \quad \tilde{\varphi} = \tilde{\varphi}_0(\mathbf{r}_\pi) + \tilde{\varphi}_1(\mathbf{r}_\pi)z + \tilde{\varphi}_2(\mathbf{r}_\pi)\frac{z^2}{2}, \tag{52}$$

where owing to (47) and (50)

$$\tilde{u}_z^1 = \mathbf{v}_1 \cdot \mathbf{r}_\pi + b_1, \qquad \tilde{u}_z^2 = \mathbf{v}_2 \cdot \mathbf{r}_\pi + b_2. \tag{53}$$

Equation (37) implies $\mathbf{u}_\pi'' = -(\mathbf{v}_1 + \mathbf{v}_2 z)/\alpha_1$ and, consequently,

$$\mathbf{u}_\pi(r, z) = \mathbf{u}_\pi^0(\mathbf{r}_\pi) + \mathbf{u}_\pi^1(\mathbf{r}_\pi) z - \frac{1}{\alpha_1}\left(\mathbf{v}_1 \frac{z^2}{2} + \mathbf{v}_2 \frac{z^3}{3!}\right). \tag{54}$$

Now from (46) and (33) we obtain $div(\mathbf{u}_\pi) = -(\alpha_2 u_z' + \beta_1 \varphi')/(\mu + \lambda)$ from which we deduce

$$div(\mathbf{u}_\pi^0) + z[div(\mathbf{u}_\pi^1)] = -\frac{1}{\mu + \lambda}\left[(\alpha_2 u_z^1 + \beta_1 \varphi_1) + z(\alpha_2 u_z^2 + \beta_1 \varphi_2)\right], \tag{55}$$

so that

$$div(\mathbf{u}_\pi^0) = -\frac{1}{\mu + \lambda}(\alpha_2 u_z^1 + \beta_1 \varphi_1), \qquad div(\mathbf{u}_\pi^1) = -\frac{1}{\mu + \lambda}(\alpha_2 u_z^2 + \beta_1 \varphi_2) \tag{56}$$

and finally by using this in (28)

$$\Delta_\pi \tilde{\varphi} = -F_2 u_z^2 - G_2 \varphi_2 \tag{57}$$

with

$$F_2 = \frac{\alpha_2(\beta_1 + \beta_2)}{2(\mu + \lambda)} + B_2, \quad G_2 = \frac{\beta_1(\beta_1 + \beta_2)}{2(\mu + \lambda)} + A_2.$$

Because of (40), (41) and (44) and using $(51)_2$ and $(52)_2$ we have

$$\Delta_\pi \tilde{\varphi}' = \Delta_\pi \tilde{\varphi}_1 + z \Delta_\pi \tilde{\varphi}_2 = 0, \qquad \Delta_\pi \tilde{u}_z' = \Delta_\pi \tilde{u}_z^1 + z \Delta_\pi \tilde{u}_z^2 = 0 \tag{58}$$

and finall

$$\Delta_\pi \tilde{\varphi}_1 = 0, \quad \Delta_\pi \tilde{\varphi}_2 = 0, \tag{59}$$
$$\Delta_\pi \tilde{\varphi}_0 = -F_2 u_z^2 - G_2 \varphi_2, \tag{60}$$
$$\Delta_\pi \tilde{u}_z^1 = 0, \quad \Delta_\pi \tilde{u}_z^2 = 0, \tag{61}$$
$$\Delta_\pi \tilde{u}_z^0 = F_3 u_z^2 - G_3 \varphi_2, \tag{62}$$

with $F_3 = 2A_1 - \alpha_1 \alpha_2/(\mu + \lambda)$ and $G_3 = 2B_1 - \alpha_1 \beta_2/(\mu + \lambda)$. We note that the functions $\tilde{u}_z^1$ and $\tilde{u}_z^2$, given in (53), satisfy Eq. (61).

To obtain the boundary condition for $\tilde{u}_z^0$ we note that from (2), (21) and (25) it follows

$$\left[\alpha_1 \mathbf{u}_\pi' \cdot \mathbf{n} + (grad(\tilde{u}_z)) \cdot \mathbf{n}\right]_{\partial C_\pi \times l} = 0 \tag{63}$$

and so, because of $(51)_2$ and (54),

$$\left[(grad(\tilde{u}_z^0)) \cdot \mathbf{n} + \alpha_1 \mathbf{u}_\pi^1 \cdot \mathbf{n}\right]_{\partial C_\pi \times l} = 0. \tag{64}$$

To determine the field $\mathbf{u}_\pi^0$ and $\mathbf{u}_\pi^1$ we use

$$Sym(grad(\mathbf{u}_\pi)) = -\frac{\alpha_2 u_z' + \beta_1 \varphi'}{2(\mu + \lambda)} \mathbf{I}_\pi$$

from which we conclude that

$$Sym(grad(\mathbf{u}_\pi^0)) = -\frac{\alpha_2 u_z^1 + \beta_1 \varphi_1}{2(\mu + \lambda)} \mathbf{I}_\pi, \quad Sym(grad(\mathbf{u}_\pi^1)) = -\frac{\alpha_2 u_z^2 + \beta_1 \varphi_2}{2(\mu + \lambda)} \mathbf{I}_\pi. \tag{65}$$

In accordance with (26) we can set

$$Sym(grad(\mathbf{u}_\pi^0)) = -\frac{\alpha_2 u_z^1 + \beta_1 \varphi_1}{2(\mu + \lambda)} \mathbf{I}_\pi, \quad Sym(grad(\mathbf{u}_\pi^1)) = -\frac{\alpha_2 u_z^2 + \beta_1 \varphi_2}{2(\mu + \lambda)} \mathbf{I}_\pi, \tag{66}$$
$$Skw(grad(\mathbf{u}_\pi^0)) = *g_1, \qquad\qquad Skw(grad(\mathbf{u}_\pi^1)) = *g_2, \tag{67}$$

where because of (26)

$$\mathrm{grad}(g_1) = -\frac{*\mathrm{grad}(\alpha_2 u_z^1 + \beta_1 \varphi_1)}{2(\mu + \lambda)}, \quad \mathrm{grad}(g_2) = -\frac{*\mathrm{grad}(\alpha_2 u_z^2 + \beta_1 \varphi_2)}{2(\mu + \lambda)}. \tag{68}$$

In this way, once the problem for the potential $\varphi$ is solved, one can calculate $g_1$ and $g_2$ and fin $\mathbf{u}_\pi^0$ and $\mathbf{u}_\pi^1$. This we will do in the following for two classes of boundary conditions,

$$\mathbf{D} \cdot \mathbf{n}|_{\partial C_\pi \times l} = 0, \quad \varphi(\mathbf{r}_\pi, z) = \varphi_0(\mathbf{r}_\pi) + z\varphi_1(\mathbf{r}_\pi) + z^2 \varphi_2(\mathbf{r}_\pi) \mathbf{r}_\pi \in \partial C_\pi.$$

## 4 Piezoelectric beam in contact with low permitivity medium: $\mathbf{D} \cdot \mathbf{n}|_{\partial C_\pi \times l} = 0$

From $\mathbf{D} \cdot \mathbf{n}|_{\partial C_\pi \times l} = 0$ the following boundary conditions emerge

$$\mathrm{grad}(\widetilde{\varphi}_0) \cdot \mathbf{n} = \frac{\beta_1 + \beta_2}{2} \mathbf{u}_\pi^1 \cdot \mathbf{n}, \tag{69}$$

$$\mathrm{grad}(\widetilde{\varphi}_1) \cdot \mathbf{n} = -\frac{\beta_1 + \beta_2}{2} \mathbf{v}_1 \cdot \mathbf{n}, \tag{70}$$

$$\mathrm{grad}(\widetilde{\varphi}_2) \cdot \mathbf{n} = -\frac{\beta_1 + \beta_2}{2} \mathbf{v}_2 \cdot \mathbf{n}, \tag{71}$$

so that, from Eqs. (59) and (61), the relations

$$\widetilde{\varphi}_1 = -\frac{\beta_1 + \beta_2}{2} \mathbf{v}_1 \cdot \mathbf{r}_\pi + \widetilde{\varphi}_1^0, \tag{72}$$

$$\widetilde{u}_z^1 = \mathbf{v}_1 \cdot \mathbf{r}_\pi + b_1, \tag{73}$$

$$\widetilde{\varphi}_2 = -\frac{\beta_1 + \beta_2}{2} \mathbf{v}_2 \cdot \mathbf{r}_\pi + \widetilde{\varphi}_2^0, \tag{74}$$

$$\widetilde{u}_z^2 = \mathbf{v}_2 \cdot \mathbf{r}_\pi + b_2 \tag{75}$$

and consequently

$$\varphi_1 = \frac{1}{\gamma_1 \alpha_1 + \beta_2(\beta_1 + \beta_2)} \left( \frac{\beta_1 + \beta_2}{2} b_1 + \alpha_1 \widetilde{\varphi}_1^0 \right), \tag{76}$$

$$u_z^1 = \frac{1}{\alpha_1} \mathbf{v}_1 \cdot \mathbf{r}_\pi + \frac{\gamma_1 b_1 - 2\beta_2 \widetilde{\varphi}_1^0}{\gamma_1 \alpha_1 + \beta_2(\beta_1 + \beta_2)}, \tag{77}$$

$$\varphi_2 = \frac{1}{\gamma_1 \alpha_1 + \beta_2(\beta_1 + \beta_2)} \left( \frac{\beta_1 + \beta_2}{2} b_2 + \alpha_1 \widetilde{\varphi}_2^0 \right), \tag{78}$$

$$u_z^2 = \frac{1}{\alpha_1} \mathbf{v}_2 \cdot \mathbf{r}_\pi + \frac{\gamma_1 b_2 - 2\beta_2 \widetilde{\varphi}_2^0}{\gamma_1 \alpha_1 + \beta_2(\beta_1 + \beta_2)} \tag{79}$$

can be derived. The compatibility condition for Neuman's problem yields two conditions

$$\int_{C_\pi} \sigma' = \int_{C_\pi} A_1 u_z^2 + B_1 \varphi_2 = 0 \quad \int_{C_\pi} D_3' = \int_{C_\pi} B_2 u_z^2 + A_2 \varphi_2 = 0. \tag{80}$$

which are equivalent to

$$b_2 = -\mathbf{v}_2 \cdot \mathbf{r}_B, \quad \widetilde{\varphi}_2^0 = \frac{\beta_1 + \beta_2}{2\alpha_1} \mathbf{v}_2 \cdot \mathbf{r}_B \tag{81}$$

with

$$\mathbf{r}_B := \frac{1}{A_{C_\pi}} \int_{C_\pi} \mathbf{r}_\pi.$$

Thus,

$$\varphi_1 = \frac{1}{\gamma_1\alpha_1 + \beta_2(\beta_1 + \beta_2)} \left(\frac{\beta_1 + \beta_2}{2} b_1 + \alpha_1 \widetilde{\varphi}_1^0\right), \quad \varphi_2 = 0, \tag{82}$$

$$u_z^1 = \frac{1}{\alpha_1}\mathbf{v}_1 \cdot \mathbf{r}_\pi + \frac{\gamma_1 b_1 - 2\beta_2 \widetilde{\varphi}_1^0}{\gamma_1\alpha_1 + \beta_2(\beta_1 + \beta_2)}, \qquad u_z^2 = \frac{1}{\alpha_1}\mathbf{v}_2 \cdot (\mathbf{r}_\pi - \mathbf{r}_B). \tag{83}$$

Equations (60) and (62) can thus finall be written as

$$\Delta_\pi \widetilde{\varphi}_0 = -\frac{F_2}{\alpha_1} \mathbf{v}_2 \cdot (\mathbf{r}_\pi - \mathbf{r}_B), \tag{84}$$

$$grad(\widetilde{\varphi}_0) \cdot \mathbf{n}|_{\partial C_\pi \times l} = \frac{\beta_1 + \beta_2}{2} \mathbf{u}_\pi^1 \cdot \mathbf{n}|_{\partial C_\pi \times l}, \tag{85}$$

$$\Delta_\pi \tilde{u}_z^0 = -\frac{F_3}{\alpha_1} \mathbf{v}_2 \cdot (\mathbf{r}_\pi - \mathbf{r}_B), \tag{86}$$

$$grad(\mathbf{u}_\pi^0) \cdot \mathbf{n}|_{\partial C_\pi \times l} = -\alpha_1 \mathbf{u}_\pi^1 \cdot \mathbf{n}|_{\partial C_\pi \times l}. \tag{87}$$

As for $\mathbf{u}_\pi^0$ and $\mathbf{u}_\pi^1$ we fin because of (26)

$$\mathbf{u}_\pi^0(\mathbf{r}_\pi) = \mathbf{u}_\pi^1(o) + \lambda_0 \mathbf{r}_\pi + \mu_1^0 * \mathbf{r}_\pi + \frac{1}{2}\left[(\lambda_1 \cdot \mathbf{r}_\pi)\mathbf{r}_\pi + (*\lambda_1 \cdot \mathbf{r}_\pi) * \mathbf{r}_\pi\right], \tag{88}$$

$$\mathbf{u}_\pi^1(\mathbf{r}_\pi) = \mathbf{u}_\pi^2(o) - \frac{\alpha_2 \mathbf{v}_2 \cdot \mathbf{r}_B \mathbf{r}_\pi}{4\alpha_1(\mu + \lambda)} + \mu_2^0 * \mathbf{r}_\pi + \tag{89}$$

$$-\frac{\alpha_2}{4\alpha_1(\mu+\lambda)}\left[(\mathbf{v}_2 \cdot (\mathbf{r}_\pi - \mathbf{r}_B))\mathbf{r}_\pi + (*\mathbf{v}_2 \cdot \mathbf{r}_\pi) * \mathbf{r}_\pi\right],$$

where

$$\lambda_1 = -\frac{\alpha_2}{2\alpha_1(\mu+\lambda)}\mathbf{v}_1, \quad \lambda_0 = -\frac{\alpha_2}{2\alpha_1(\mu+\lambda)}\left[b_1 + \widetilde{\varphi}_1^0 \frac{(\alpha_1\beta_1 - 2\alpha_2\beta_2)}{\gamma_1\alpha_1 + \beta_2(\beta_1+\beta_2)}\right];$$

$\mu_1^0$ and $\mu_2^0$ are integration constants.

In this way the problem is completely solved in terms of twelve arbitrary constants. As $\mathbf{u}_\pi^0(o)$, $\mathbf{u}_\pi^1(o)$ and $\mu_0^1$ correspond to rigid motions that shall henceforth be disregarded and thus omitted, there remain seven constants, six corresponding to the usual SV solution and the seventh related to the flu of the electric displacement $\mathbf{D}$ through the bases. In particular we note that by assuming all the constants to vanish but $\widetilde{\varphi}_1^0$ and $b_1$, we f nd

$$\varphi_2 = 0, \quad u_z^2 = 0, \quad \mathbf{u}_\pi^1 = \mathbf{0}, \quad u_z^0 = 0, \quad \varphi_0 = 0, \tag{90}$$

$$\varphi_1 = \frac{1}{\gamma_1\alpha_1 + \beta_2(\beta_1+\beta_2)}\left(\frac{\beta_1+\beta_2}{2}b_1 + \alpha_1\widetilde{\varphi}_1^0\right), \tag{91}$$

$$u_z^1 = \frac{\gamma_1 b_1 - 2\beta_2 \widetilde{\varphi}_1^0}{\gamma_1\alpha_1 + \beta_2(\beta_1+\beta_2)}, \tag{92}$$

$$\mathbf{u}_\pi^0 = -\frac{\alpha_2}{2\alpha_1(\mu+\lambda)}\left[b_1 + \widetilde{\varphi}_1^0 \frac{(\alpha_1\beta_1 - 2\alpha_2\beta_2)}{\gamma_1\alpha_1 + \beta_2(\beta_1+\beta_2)}\right]\mathbf{r}_\pi, \tag{93}$$

$$V = l\varphi_1. \tag{94}$$

$V$ is the potential difference between the bases. Consequently the resultant force and fiel are given by

$$\int_{C_\pi \times 0} \sigma = \int_{C_\pi \times 0} A_1 u_z' + B_1 \varphi'$$

$$= A\left\{\frac{\widetilde{\varphi}_1^0(-2A_1\beta_2 + B_1\alpha_1) + b_1(A_1\gamma_1 + B_1(\beta_1+\beta_2))}{\gamma_1\alpha_1 + \beta_2(\beta_1+\beta_2)}\right\}, \tag{95}$$

$$\int_{C_\pi \times 0} D_3 = \int_{C_\pi \times 0} B_2 u_z' + A_2 \varphi'$$

$$= A\left\{\frac{\widetilde{\varphi}_1^0(-2B_2\beta_2 + A_2\alpha_1) + b_1(B_2\gamma_1 + A_2(\beta_1+\beta_2))}{\gamma_1\alpha_1 + \beta_2(\beta_1+\beta_2)}\right\}, \tag{96}$$

where $A$ is the area of the cross section.

Because of (96) and (93) we see that an electric fiel applied through the bases of the SV cylinder produces an elongation whose Poisson-effect is controlled. In fact, choosing

$$\tilde{\varphi}_1^0 = b_1 \frac{\gamma_1 \alpha_1 + \beta_2(\beta_1 + \beta_2)}{2\alpha_2 \beta_2 - \alpha_1 \beta_1} \tag{97}$$

we fin $\lambda_0 = 0$ and consequently (93)

$$\mathbf{u}_\pi^0 = \mathbf{0}. \tag{98}$$

This means that if we apply, between the bases, the potential difference $V$ given by (94) in which $\tilde{\varphi}_1^0$ is given by (97), then the Poisson effect in extension vanishes.

We postpone the discussion of other deformation cases to a forthcoming investigation. We simply remark here that the only change with respect to the standard SV problem -when no transversal electric fiel is applied- concerns the elasticity constants. More interesting results are likely to arise in the deformation of piezoelectric cylinders embedded in an external electric field as is suggested by the results obtained in the following section.

## 5 Almansi type boundary condition:
$\varphi|_{\partial C_\pi \times I} = \varphi_0(r_\pi) + z\varphi_1(r_\pi) + z^2 \varphi_2(r_\pi), \quad r_\pi \in \partial C_\pi$

*5.1 The general elliptic problem*

In the considered case the following Dirichlet problem for the potential must hold:

$$\Delta_\pi \varphi_1 = 0 \qquad \text{in } C_\pi, \qquad \varphi_1|_{\partial C_\pi \times I} = \varphi_1^M, \tag{99}$$

$$\Delta_\pi \varphi_2 = 0 \qquad \text{in } C_\pi, \qquad \varphi_2|_{\partial C_\pi \times I} = \varphi_2^M, \tag{100}$$

$$\Delta_\pi \varphi_0 = Z_2 \tilde{u}_z^2 + Z_3 \varphi_2 \qquad \text{in } C_\pi, \qquad \varphi_0|_{\partial C_\pi \times I} = \varphi_0^M, \tag{101}$$

with

$$Z_2 = -\frac{1}{\alpha_1} \left( \frac{\alpha_1 F_2 + F_3(\beta_1 + \beta_2)/2}{\alpha_1 \gamma_1 + \beta_2(\beta_1 + \beta_2)} \right),$$

$$Z_3 = -\frac{\alpha_1 G_2 + G_3(\beta_1 + \beta_2)/2}{\alpha_1 \gamma_1 + \beta_2(\beta_1 + \beta_2)} - \frac{2\beta_2}{\alpha_1} \left( \frac{\alpha_1 F_2 + F_3(\beta_1 + \beta_2)/2}{\alpha_1 \gamma_1 + \beta_2(\beta_1 + \beta_2)} \right).$$

Once the $\varphi_2$-problem is solved we can fin $\mathbf{u}_\pi^1$ (see (26)) by solving

$$Sym(grad(\mathbf{u}_\pi^1)) = \left( Z_4 \tilde{u}_z^2 + Z_5 \varphi_2 \right) \mathbf{I}_\pi. \tag{102}$$

Similarly, $\mathbf{u}_\pi^0$ follows from

$$Sym(grad(\mathbf{u}_\pi^0)) = \left( Z_4 \tilde{u}_z^1 + Z_5 \varphi_1 \right) \mathbf{I}_\pi \tag{103}$$

once $\varphi_1$ is known. Thus we have a well posed problem for $\tilde{u}_z^0$

$$\Delta_\pi \tilde{u}_z^0 = Z_0 \tilde{u}_z^2 + Z_1 \varphi_2, \quad grad\left(\tilde{u}_z^0\right) \cdot \mathbf{n}|_{\partial C_\pi \times I} = \alpha_1 \mathbf{u}_\pi^1 \cdot \mathbf{n}|_{\partial C_\pi \times I}, \tag{104}$$

where

$$Z_0 = -\frac{F_3}{\alpha_1}, \quad Z_1 = \frac{2\beta_2}{\alpha_1} - \frac{\alpha_1 G_2 + G_3(\beta_1 + \beta_2)/2}{\alpha_1 \gamma_1 + \beta_2(\beta_1 + \beta_2)}.$$

*5.2 A boundary value problem for a SV cylinder of rectangular cross section*

Let us apply these formulas to a SV cylinder of rectangular section $[0, x_0] \times [0, y_0]$. Assuming the boundary conditions

$$\varphi_2(x, 0) = k_0, \quad \varphi_2(x, y_0) = k_1, \qquad x \in [0, x_0], \tag{105}$$

$$\varphi_2(0, y) = k_0 + \frac{k_1 - k_0}{y_0} y = \varphi_2(x_0, y), \quad y \in [0, y_0], \tag{106}$$

for $\varphi_2$, we arrive at

$$\varphi_2(\mathbf{r}_\pi) = k_0 + \mathbf{k} \cdot \mathbf{r}_\pi, \text{ with } \mathbf{k} = (0, (k_1 - k_0)/y_0). \tag{107}$$

The condition $\tau \cdot \mathbf{n}|_{\partial C_\pi \times l} = 0$ allows $b_2$ to be determined, viz.,

$$b_2 = \mathbf{v}_2 \cdot \mathbf{r}_B + (k_0 + \mathbf{k} \cdot \mathbf{r}_B)\left(2\beta_2 - \frac{B_1 \alpha_1}{A_1}\right), \tag{108}$$

from which

$$\tilde{u}_z^2 = \mathbf{v}_2 \cdot (\mathbf{r}_\pi - \mathbf{r}_B) + (k_0 + \mathbf{k} \cdot \mathbf{r}_B)\left(2\beta_2 - \frac{B_1 \alpha_1}{A_1}\right). \tag{109}$$

In conclusion, we have (see $(66)_2$, $(67)_2$ and (26)) [recall the definitio of the * operator]

$$\mathbf{u}_\pi^1 = \frac{1}{2}[(\Omega_0 \cdot (\mathbf{r}_\pi - \mathbf{r}_B))\mathbf{r}_\pi + (*\Omega_0 \cdot \mathbf{r}_\pi) * \mathbf{r}_\pi] + \overline{\Omega}_1 \mathbf{r}_\pi + \mu_1^0 * \mathbf{r}_\pi + \mathbf{u}_\pi^1(o) \tag{110}$$

with

$$\Omega_0 = Z_4 \mathbf{v}_2 + Z_5 \mathbf{k}, \quad \overline{\Omega}_1 = (\mathbf{k} \cdot \mathbf{r}_B + k_0)\left[Z_5 + Z_4\left(2\beta_2 - \frac{B_1 \alpha_1}{A_1}\right)\right],$$

$$Z_4 = -\frac{\alpha_2}{2\alpha_1(\mu + \lambda)}, \quad Z_5 = \frac{2\beta_2 \alpha_2 - \alpha_1 \beta_1}{2\alpha_1(\mu + \lambda)}.$$

For $\tilde{u}_z^0$ the boundary value problem reads

$$\Delta_\pi \tilde{u}_z^0 = \Omega_2 \cdot (\mathbf{r}_\pi - \mathbf{r}_B) + \overline{\Omega}_3, \quad \text{grad}(\tilde{u}_z^0) \cdot \mathbf{n}|_{\partial C_\pi \times l} = -\alpha_1 \mathbf{u}_\pi^1 \cdot \mathbf{n}|_{\partial C_\pi \times l} \tag{111}$$

with

$$\Omega_2 = Z_0 \mathbf{v}_2 + Z_1 \mathbf{k}, \quad \overline{\Omega}_3 = (\mathbf{k} \cdot \mathbf{r}_B + k_0)\left[Z_1 + Z_0\left(2\beta_2 - \frac{B_1 \alpha_1}{A_1}\right)\right].$$

The evaluation of the resultant forces on the bases of the cylinder in terms of the kinematic parameters introduced is similar to what was done in [10]. We remark here that for flexur the resultant shear force takes the form

$$\mathbf{q} := \int_{C_\pi} \mathbf{r}_\pi \sigma', \tag{112}$$

in which

$$\sigma' = \mathbf{k} \cdot (\mathbf{r}_\pi - \mathbf{r}_B)\left(B_1 - \frac{2\beta_2 A_1}{al_1}\right) + \frac{A_1}{\alpha_1} \mathbf{v}_2 \cdot (\mathbf{r}_\pi - \mathbf{r}_B),$$

so that $\mathbf{q}$ can be written as

$$\mathbf{q} = *\mathbf{J}_B(Y\mathbf{v}_2 + \overline{Y}\mathbf{k}) \tag{113}$$

where

$$Y = \frac{A_1}{\alpha_1}, \quad \overline{Y} = B_1 - \frac{2\beta_2 A_1}{\alpha_1}, \quad \mathbf{J}_B := \int (\mathbf{r}_\pi - \mathbf{r}_B) \otimes (\mathbf{r}_\pi - \mathbf{r}_B) dA;$$

$\mathbf{J}_B$ is the Euler tensor of inertia.

The above equations make evident that by fixing the resultant shear force, **q**, the value of $\mathbf{v}_2$ can be controlled by varying **k**. Moreover, because $u_z^0 := (\tilde{u}_z^0 - 2\beta_2 \varphi_0)/\alpha_1$ warping can be controlled by correspondingly controlling the potential $\varphi_0$. Finally, once the problem for $\varphi_1$ is solved, we can calculate $\mathbf{u}_\pi^0$ and consequently derive the full expression for $\mathbf{u}_\pi$

$$\mathbf{u}_\pi(r,z) = \mathbf{u}_\pi^0(\mathbf{r}_\pi) + \mathbf{u}_\pi^1(\mathbf{r}_\pi)z - \frac{1}{\alpha_1}\left(\mathbf{v}_1 \frac{z^2}{2} + \mathbf{v}_2 \frac{z^3}{3!}\right). \tag{114}$$

We postpone the discussion of how the deformation of SV cylinders can be controlled by applying the appropriate electric field to further investigations.

## 6 Conclusions

In the literature (see [7] for a detailled discussion of the results already available) all possible coupling effects between electric and mechanic phenomena present in the SV problem have been investigated. In [2] the SV principle is proved for linear piezoelectric materials, while in [1, 3] an exhaustive study of TIP constitutive equations is performed. These papers implicitly suggest that more interesting coupling effects must occur when (at least for the electric potential) *Almansi-type* boundary conditions are considered.

In the present paper we proved that by simply considering quadratic Almansi-type boundary conditions for the electric potential it is possible -in SV cylinders which are not loaded on the lateral wall- to significantly influence by means of the externally applied electric potential the displacement field that is induced by shear forces applied on its bases. In particular we proved that it is possible to control

 i) the Poisson effect-in SV deformation of a piezoelectric cylinder with lateral boundary unloaded and in contact with a relatively low permittivity medium-by applying a potential difference between its bases;
 ii) the deformation of axes of a particular SV piezoelectric cylinder in flexure when its lateral boundary is not loaded but it is in contact with a family of conductors whose potential varies quadratically with the $z$-coordinate.

While the first effect has been partially studied in the literature, we could not find anywhere a description of the second.

We underline that the set of Eqs. (99–104) represents an interesting elliptic problem which, to our knowledge, is not studied in the literature. In our opinion it is worth further investigation with a view to applications.

*Acknowledgements.* This research was partially supported by Italian MURST. The authors thank professor R.C. Batra for his friendly criticism.